\documentclass[12pt,a4paper,final]{iopart}

\usepackage{iopams}  
\usepackage{graphicx}
\usepackage{cite}
\begin{document}

\title{Electron counting in a silicon single-electron pump}%

\author{Tuomo Tanttu$^1$, Alessandro Rossi$^2$, Kuan Yen Tan$^1$, Kukka-Emilia Huhtinen$^1$, Kok Wai Chan$^2$\footnote{Present address: Centre for Advanced 2D Materials and Graphene Research Centre, National University of Singapore.}, Mikko M\"ott\"onen$^1$, Andrew S. Dzurak$^2$ }%
\address{$^1$QCD Labs, COMP Centre of Excellence, Department of Applied Physics, Aalto University, P.O. Box 13500, 00076 Aalto, Finland}
\address{$^2$School of Electrical Engineering \& Telecommunications, The University of New South Wales, Sydney 2052, Australia}
\ead{tuomo.tanttu@aalto.fi}

\begin{abstract}
We report electron counting experiments in a silicon metal--oxide--semiconductor quantum dot architecture which has been previously demonstrated to generate a quantized current in excess of 80~pA with uncertainty below 30~parts per million. Single-shot detection of electrons pumped into a reservoir dot is performed using a capacitively coupled single-electron transistor. We extract the full probability distribution of the transfer of $n$ electrons per pumping cycle for $n=0,1,2,3,\textrm{ and }4$. We find that the probabilities extracted from the counting experiment are in agreement with direct current measurements in a broad range of dc electrochemical potentials of the pump. The electron counting technique is also used to confirm the improving robustness of the pumping mechanism with increasing electrostatic confinement of the quantum dot

\end{abstract}

\vspace{2pc}
\submitto{\NJP}

\section{Introduction}

Recent development in the field of single-charge pumping has provided a basis for the emerging quantum standard of the ampere in the International System of Units (SI)\cite{Pekola2013}. This standard will be based on an agreed value for the elementary charge $e$ and the frequency $f$, the product of which yields the ampere.

Single-charge pumps and turnstiles have been implemented in many different physical systems including normal-metal tunnel junction devices \cite{Pothier1992, Keller1996, Keller1999}, superconducting devices \cite{Geerligs1991,Vartiainen2007, Mottonen2008}, hybrid superconductor--normal-metal turnstiles \cite{Pekola2008, Maisi2011}, semiconductor quantum dots \cite{Kouwenhoven1991,Blumenthal2007,Chan2011,Jehl2012,Rossi2014, Connolly2013}, and single atom-sized impurities \cite{Lansbergen2012,Roche2012,Yamahata2014,Tettamanzi2014}. A satisfactory relative pumping accuracy at the $10^{-8}$ level has only been demonstrated in normal-metal devices in the picoampere range \cite{Keller1996}. This current, however, falls significantly below~100~pA which is required for a practical realization of the quantum current standard \cite{Feltin2009}. The most accurate single-electron pumps that produce high enough current are thus far based on GaAs quantum dots \cite{Giblin2012}. Recently, an uncertainty 0.2 parts per million~(ppm) levels has been reached at 87-pA current \cite{Stein2015}.

Silicon quantum dots~\cite{Fujiwara2001,Fujiwara2004,Fujiwara2008,Chan2011,Rossi2014} 
provide a promising alternative to the GaAs platform. Devices fully based on silicon have exhibited greatly suppressed $1/f$ noise and absence of large amplitude background charge jumps \cite{Koppinen2013}. To date the most accurate silicon single-electron pumps produce a pumped current of~80~pA with uncertainty below~30~ppm \cite{Rossi2014}. 

The accuracy of the electron pump is essentially given by missed or excess electrons pumped per cycle. It is possible to arrange the electron pumps such that the pumping errors can be in-situ observed with a nearby charge sensor, thus providing a self-referenced current source. Although several experiments \cite{Keller1996, Keller2000error, Jehl2003, Kautz2000, Kautz1999, Fujiwara2001, Nishiguchi2006, Yamahata2011, Fricke2013, Fricke2014, Yamahata2014,Yamahata2014_revB} provide observations on the pumping errors and the number of electrons transferred per cycle, a thorough comparison of the direct current provided by the electron pump and the results of the electron counting scheme is lacking. Only comparison between electron counting and the current flowing through a non-driven system has been reported \cite{Bylander2005}.

In this paper, we demonstrate electron counting in a silicon electron pump utilizing the quantum dot architecture which has provided the most accurate results in silicon \cite{Rossi2014}, thus providing a proof of concept for a self-referenced silicon charge pump. Furthermore, the average number of pumped electrons per cycle, $n$, extracted from our electron counting scheme agrees with that obtained from the pumped direct current. This result verifies the consistency between these two schemes.

\section{Experimental methods}

\begin{figure}
\begin{center}
\includegraphics[width=14cm]{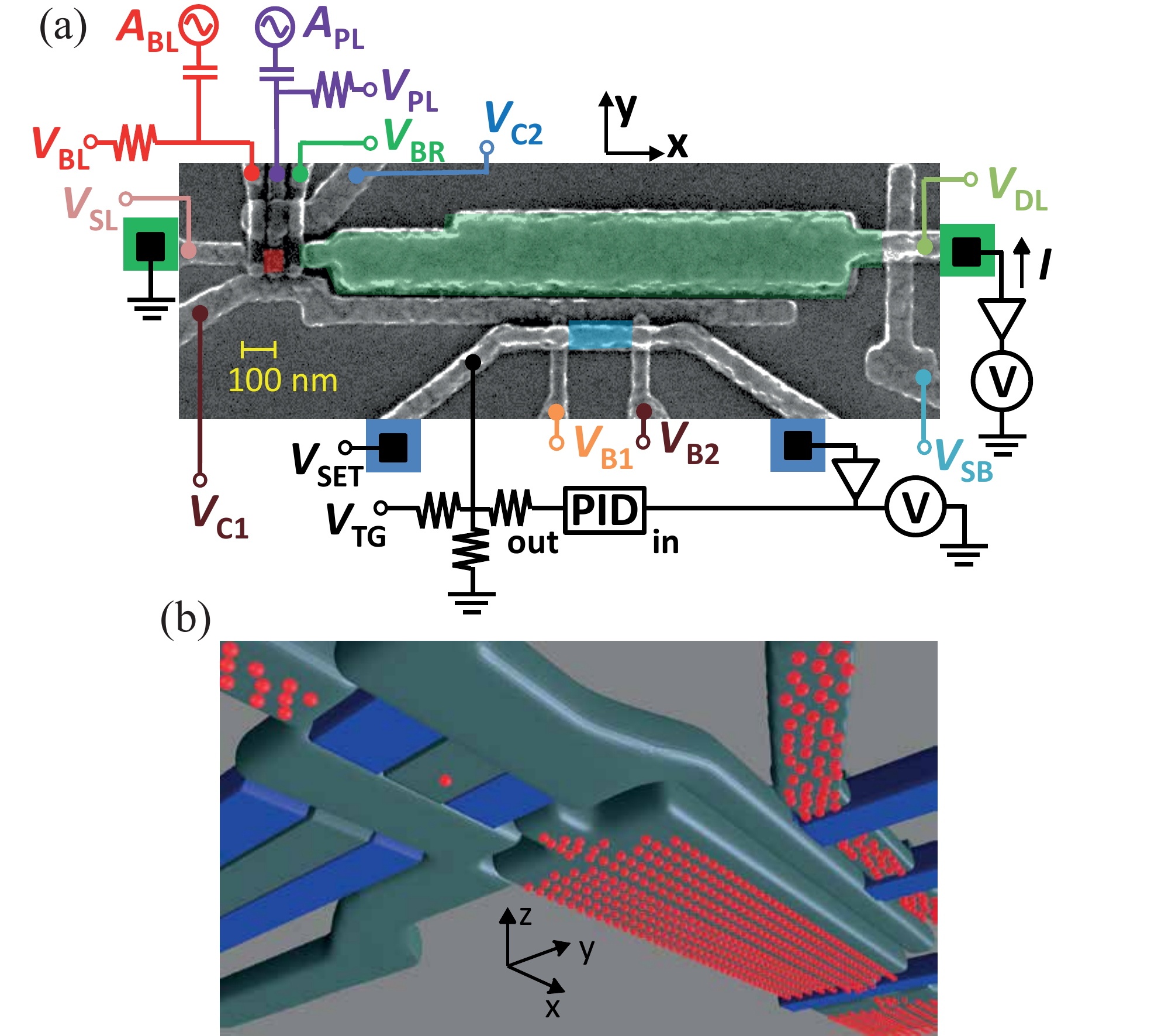}
\caption{(a) False-color scanning electron micrograph of a device similar to the one used in the experiments together with a sketch of the measurement setup. The quantum dot~(QD) used to pump electrons is highlighted in red. The reservoir dot (RES), into which the electrons are pumped, is green and the dot of the single-electron transistor sensor is highlighted in blue. The green (blue) squares represent the source and drain ohmic contacts of the pump (sensor). The gates are labelled according to their indicated dc voltages. (b) Schematic illustration of the device gate layout from below. Electric fields induced by the gate voltages are used to selectively accumulate electrons (red spheres) at the Si/SiO$_2$ interface (not shown). Gates highlighted in blue are used to form tunnel barriers.}
\label{measurement_setup}
\end{center}
\end{figure}

\begin{figure}
\begin{center}
\includegraphics[width=8cm]{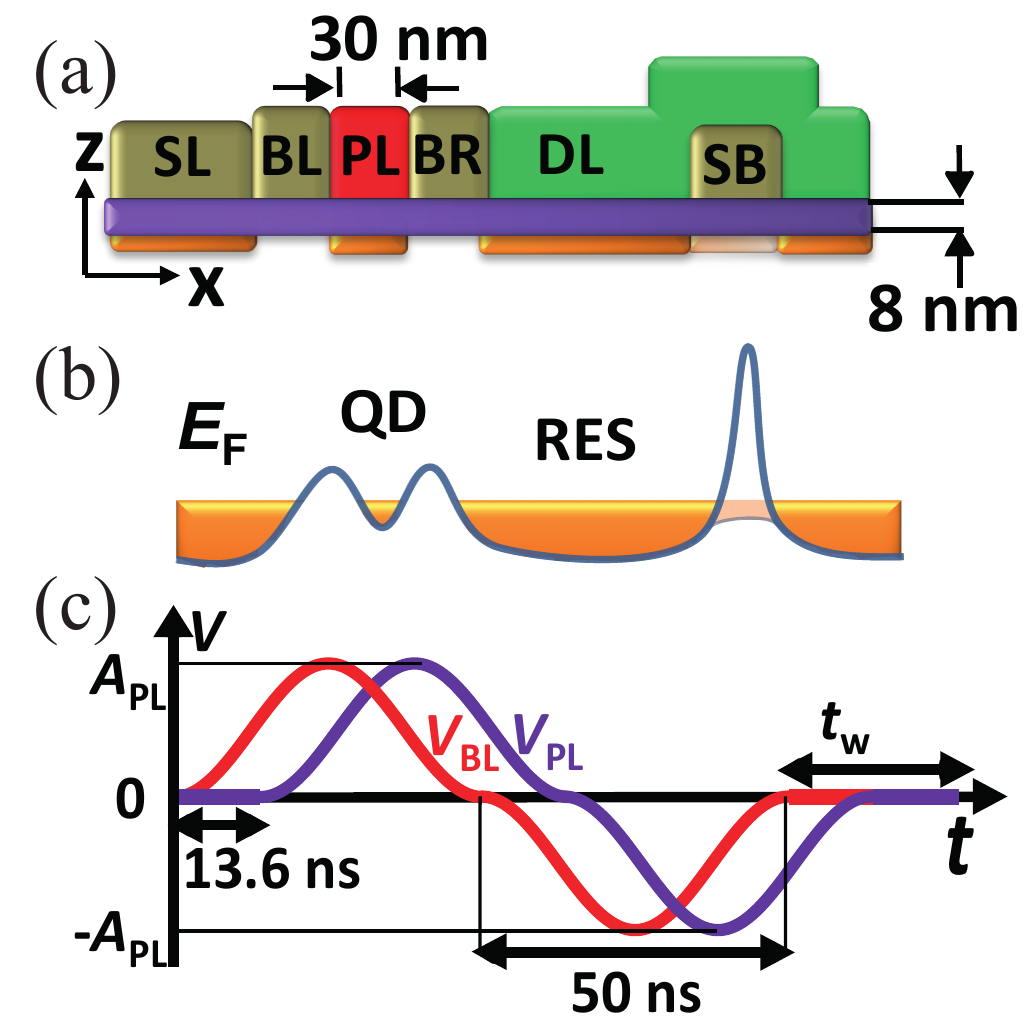}
\caption{(a) Schematic cross section and (b) potential landscape of the device along~$x$. By changing~$V_\textrm{SB}$ we can vary the reservoir to be either a large dot (electron counting scheme) or a current lead (direct-current scheme). The 2DEG is indicated with orange. The shaded area indicates 2DEG which can be induced or depleted by varying~$V_\textrm{SB}$. (c) The repeated waveforms for the voltage drives on BL (red) and PL (purple) used in the experiment. }
\label{measurement_setup2}
\end{center}
\end{figure}

Our device shown in Figs.~\ref{measurement_setup}(a), ~\ref{measurement_setup}(b), and~\ref{measurement_setup2}(a) is fabricated using metal--oxide--semiconductor~(MOS) technology on a near-intrinsic silicon substrate with 8-nm thermally grown SiO$_2$ gate-oxide \cite{Rossi2014,Rossi2015}. The aluminum gates are defined with electron beam lithography in three layers isolated from each other by thermally grown Al$_y$O$_x$. The topmost layer of gates is used to accumulate a two-dimensional electron gas (2DEG) at the Si/SiO$_2$ interface and the two bottom layers are used to control the electrostatic confinement of the dot in the planar directions by locally depleting the 2DEG and forming tunnel barriers. A schematic potential landscape of the device is presented in~Fig.~\ref{measurement_setup2}(b).

We employ two different measurement schemes: the \emph{direct-current scheme} and the \emph{electron counting scheme}. In the direct-current scheme, we induced a 2DEG below the source lead (SL), drain lead (DL), and switch barrier (SB) gates [see Fig.~\ref{measurement_setup}(a)]. The pump dot is induced with the plunger gate (PL) such that the left barrier (BL) and right barrier (BR) gates are used to define tunable tunnel barriers between the leads and the dot. The confining gates (C1 and C2) are set to negative voltage to tighten the dot potential as first demonstrated in Ref.~\cite{Rossi2014}. Experiments in both schemes are carried out in a cryostat with a bath temperature of 180 mK. 

The gates PL and BL are also connected to an arbitrary-waveform generator providing the voltage drive for the dot to pump the electrons from the source to the drain. As shown in Fig.~\ref{measurement_setup2}(c) the waveforms of the pulses consist of three consecutive parts: (i) voltage $s_1(t)=A_\textrm{PL/BL}[1-\cos(2\pi t/T)]/2$ for $0\leq t<T$, (ii) voltage $s_2(t)=-s_1(2T-t)$ for $T\leq t<2T$, and (iii) zero voltage for $2T\leq t\leq 2T+t_\textrm{w}$. The period of the sinusoidal part is fixed at $T=50$~ns and the pumping frequency $f=1/(2T+t_\textrm{w})$ is adjusted by changing the wait time $t_\textrm{w}\gg T$. The temporal offset of the pulses in PL and BL is~13.6~ns and the voltage amplitudes at the sample are denoted by $A_\textrm{PL}$ and $A_\textrm{BL}$, respectively. The induced current is measured from the drain side using a room-temperature transimpedance amplifier. In the direct-current scheme, we have $t_\textrm{w} = 1.9$~$\mu$s that yields~$ef = 0.08$~pA. The waveform has to be adjusted such that the integral of the positive and negative area vanishes. Otherwise we need to adjust the dc bias of the gates for each $t_w$ to achieve the desired potential due to the loss of the dc component of the waveform in the capacitor of the bias tee.

The electron counting scheme has the following differences from the scheme described above: We use a much lower $V_\textrm{SB}$ to define a reservoir dot below the DL gate bounded by SB, C1, and BR gates. The charge state of the reservoir is monitored with a capacitively-coupled single-electron transistor (SET). The SET is induced with the top gate~(TG) and barrier gates (B1 and B2). The hold time of the charge state of the reservoir was measured at gate voltages similar to the one used for the counting experiments and showed stability of several hours. 
The current through the voltage biased SET is transimpedance amplified and channeled to a proportional-integral-derivative (PID) controller which keeps the operation point of the SET fixed by compensating $V_\textrm{TG}$. Electrons are pumped to the reservoir with an identical waveform as in the direct-current scheme but with relatively long wait time $t_\textrm{w}=750$~ms. After a fixed number of subsequent pumping pulses, the reservoir is initialized by inducing a 2DEG below BL, PL, and BR so that the excess electrons flow from the reservoir back to the source.

\section{Results}

\begin{figure}
\centering
\includegraphics[width=12cm]{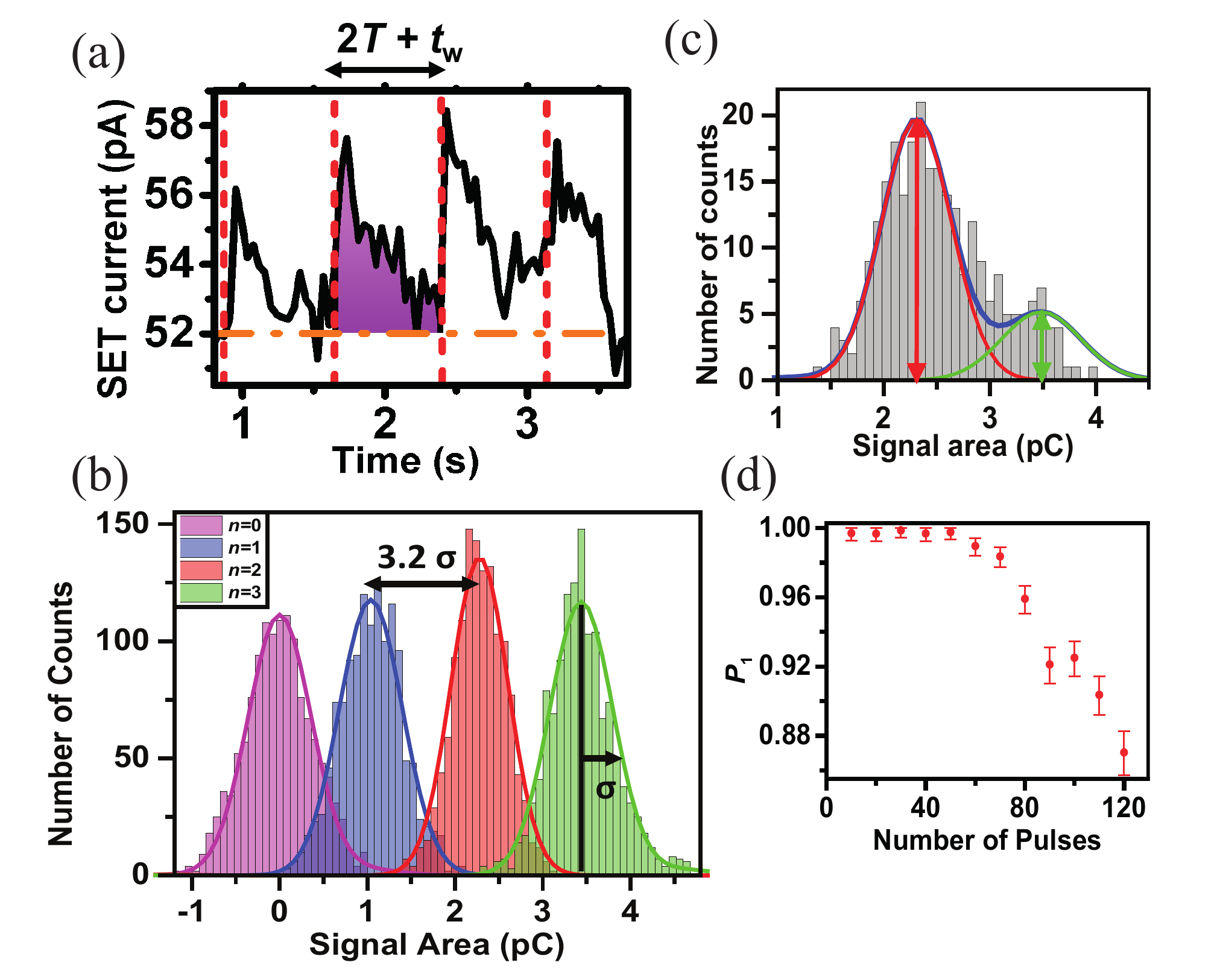}
\caption{(a) Representative trace of the SET signal in the electron counting experiments. The red vertical dashed lines indicate the time instants of the pumping events. The orange horizontal dashed line is the set point of the PID controller. The number of electrons transferred during the pumping cycle is estimated from the shaded area enclosed between the set point and the SET signal. (b) Histograms of signal areas at $V_{\textrm{PL}}$ values for which the transport is quantized at $n=0,1, 2,$~and~$3$ in the direct-current experiment. Gaussian fits are shown for each data set. (c) Histogram of SET signal areas at $V_{\textrm{PL}}=0.62$~V with 264 pumping pulses in total. The amplitudes of fitted Gaussian distributions (red and green arrows) yield the probabilities, $P_n$, as described in the text. (d) Probability of a single electron transition as a function of the number of consecutive pumping pulses with error bars indicating~95\% confidence interval. Each point is statistically evaluated as an aggregate of 2000 counts. Gate voltages are set to the values corresponding to one transferred electron per cycle in the direct-current scheme. }
\label{data_fig}
\end{figure}

Figure~\ref{data_fig}(a) shows a representative trace of the SET current signal as a function of time when electrons are pumped into the reservoir. At each pumping event, there is a clear peak in the signal which subsequently saturates back to the set point of the PID controller. The PID controller is employed to enhance the signal to noise ratio compared with the current threshold method used in other electron counting experiments \cite{Yamahata2014,Fricke2013,Fricke2014}. The advantages of this method \cite{Keller1996} are that the low-frequency noise is filtered out and the sensor works at its most sensitive operation point at all times. We characterize the electron transfers by evaluating the area between the SET current trace and the set point, as indicated in Fig.~\ref{data_fig}(a). The tunable parameters of the PID controller define the observed decay time.

In Fig.~\ref{data_fig}(b), we show histograms of the SET signal area at plunger voltages corresponding to the maximum probability of achieving $n=0,1,2,$ or 3 electrons transferred per cycle. In order to evaluate the transfer probabilities $P_n$ at different $V_\textrm{PL}$, we fit the histograms with a function $f(x) = \sum_{n}A_ng(x,b_n,\sigma_n)$, where $g(x,b_n,\sigma_n)$ is a Gaussian distribution with mean $b_n$ and standard deviation $\sigma_n$. Since the mean and standard deviation of each distribution are essentially independent of the plunger gate voltage, we determine their values using the whole data set acquired for all different voltage ranges. For the mean values we obtain $b_n = n\times 1.15$~pC and for the standard deviations $\sigma_n = 0.37$~pC for~$n\neq 2$ and $\sigma_2 = 0.33$~pC. The probabilities $P_n$ for $n$ transferred electrons are extracted using the amplitudes $A_n$ as fitting parameters for each $V_\textrm{PL}$ and computing~$P_n = A_n/\sum_{j=0}^4A_j$. In~Fig.~\ref{data_fig}(c), a representative fit at~$V_{\textrm{PL}}=0.62$~V is presented.



Figure~\ref{data_fig}(d) shows the probability of a single-electron transition as a function of the number of consecutively applied pumping pulses since the initialization of the reservoir. The error bars indicate 95\% confidence interval obtained by taking into account two error sources independently: uncertainty related to the fit of the amplitude $A_n$ and the one obtained from the Wilson score interval method. Each data point is derived from combined statistics of 2000 pulses and 200 reset events. The data show that we may inject up to 50 electrons into the reservoir without changing the probability more than 1\%. This probability decreases with increasing number of pumped electrons into the reservoir due to its increasing electrochemical potential~\cite{Jehl2003}. We estimate the capacitance between DL and the reservoir by assuming them to be parallel plate capacitors:~$C_\textrm{RES-DL}=\epsilon A/d\approx 1.8\times 10^{-15}$~F, where $\epsilon$ is the permittivity of SiO$_2$, $A$ is the area of the reservoir dot, and~$d$ is the thickness of the SiO$_2$. Thus the charging energy of the reservoir is roughly $E_C = e^2/C_\textrm{RES-DL}\approx 87$~$\mu$eV leading to a potential difference of the reservoir due to 50 excess electrons in the island of $\Delta_\textrm{RES}\approx 4.3$~mV.


\begin{figure}
\centering
\includegraphics[width=12cm]{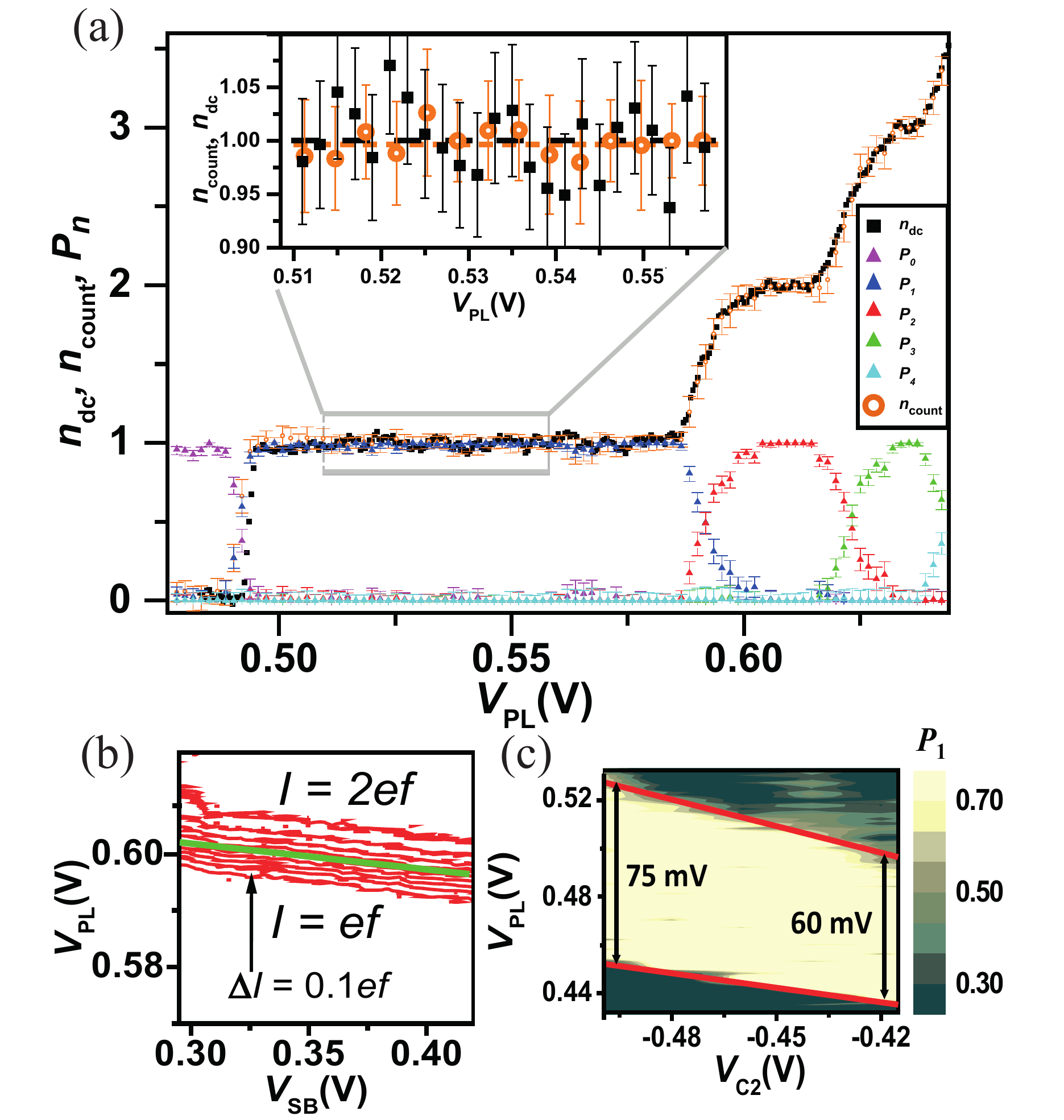}
\caption{(a) Average number of electrons pumped per cycle~$n_\textrm{dc}=I/ef$ measured in the direct-current scheme with $ef\approx 80$~fA, and probabilities $P_n$ of number of electrons pumped per cycle determined from the electron counting scheme as well as $n_\textrm{count}=\sum_n nP_n$ as functions of $V_\textrm{PL}$. The error bars indicate~95\% confidence interval. The gate voltages in the two experiments are the following: $V_\textrm{BL}=0.63$~V, $V_\textrm{BR}=0.48$~V, $V_\textrm{C2}=-1.0$~V, $V_\textrm{C1}=-0.25$~V, $V_\textrm{SL}=2.4$~V, $V_\textrm{DL}=1.9$~V, $V_\textrm{B1}=0.85$~V, and $V_\textrm{B2}=0.69$~V. In the direct-current scheme we employ $V_\textrm{SB}=0.20$~V and $V_\textrm{TG}=0.98$~V, and in the counting scheme $V_\textrm{SB}=0.39$~V and $V_\textrm{TG}=0.95$~V. The probability traces have been shifted by~$\Delta V_\textrm{PL} =-7.0$~mV to account for different values of $V_\textrm{SB}$ in the two experiments. The peak amplitudes of the rf drives are $A_\textrm{PL}=A_\textrm{BL}=0.15$~V in both cases.  Inset: Every second data point of $n_\textrm{count}$ and every fourth data point of $n_{\textrm{dc}}$ from the main panel in the voltage region highlighted by the grey rectangle. The error bars indicate~95\% confidence interval. The dashed black and orange lines represent the mean of 100 data points for $\bar{n}_\textrm{dc}$ and of 30 data points for $\bar{n}_\textrm{count}$, respectively, measured within the shown $V_\textrm{PL}$ range. (b)~Pumped direct current as a function of $V_\textrm{PL}$ and the switch barrier voltage. The other parameter values are identical to those in panel~(a). The spacing in current between the red contours is 0.1$ef$. The green line is a guide for the eye to indicate the applied linear compensation in $V_\textrm{PL}$ due to the different values of $V_\textrm{SB}$. (c) $P_1$ as a function of $V_\textrm{PL}$ and the confining gate voltage $V_\textrm{C2}$. The other parameter values are as in~(a) except for $V_\textrm{BL}=0.60$~V and $V_\textrm{TG}=1.2$~V. }
\label{measured_plateau}
\end{figure}

In order to extract the probabilities $P_n$ for $n=0,1,2,3,$~and~4 as a function of plunger voltage, we apply a reset event followed by 22 consecutive pumping pulses. We repeat this procedure 12 times for each voltage value. Based on the data presented in Fig.~\ref{data_fig}(d) where $P_1>99\%$ for up to 50 consecutive pumping pulses, we note that the choice of 22 pulses between each reset should not lead to observable underpumping for $n=1$ within the uncertainty of the counting scheme. The average number of electrons pumped per cycle can be computed from the individual probabilities as $n_\textrm{count} = \sum_nnP_n$. In Fig.~\ref{measured_plateau}(a) the probabilities $P_n$ are shown as well as $n_\textrm{count}$. The error bars are computed the same way as in Fig.~\ref{data_fig}(d). The data indicate that, by adjusting the potential of the dot,  it is possible to transfer with a single pulse up to 3 electrons with over 99\% probability. However, single electron transfers are clearly more robust than multiple electron transfers since $P_1$ is insensitive to variations of $V_\textrm{PL}$ in a significantly larger range than that of $P_2$ and $P_3$.

Fig.~\ref{measured_plateau}(a) also shows the average number of electrons transferred per cycle measured with the direct-current scheme $n_\textrm{dc}=I/ef$. Interestingly, these data are in good agreement with the counting method. Note that the curves for $P_n$ are shifted by~$-7.00$~mV in~$V_\textrm{PL}$ justified by the capacitive coupling between the SB gate and the pump dot and the fact that we need to use a different gate voltage in the direct-current scheme ($V_\textrm{SB}=0.39$~V) compared with the electron counting scheme ($V_\textrm{SB}=0.20$~V). We verified that the magnitude of the applied shift is in agreement with the observed shift of the current plateaux in the direct-current scheme [see Fig.~\ref{measured_plateau}(b)]. The electron channel under the switch barrier turns off completely around $V_\textrm{SB}=0.30$~V which prevented us from measuring the shift in this scheme at lower voltages. We neglect the shift of the plateaux due to different $V_\textrm{TG}$ used in the two schemes since it is much smaller than the shift due to~$V_\textrm{SB}$. 

In the electron counting scheme, the rising edge to the first plateau shifts in $V_\textrm{PL}$ as a function of number of excess electrons in the reservoir. Since we average over 22 pumped electrons this shift broadens the rise to the first plateau in Fig.~\ref{measured_plateau}(a). This effect is not clearly visible for the other, notably broad, steps.

In the inset of~Fig.~\ref{measured_plateau}(a), the quantized electron pumping at the $n=1$ plateau is compared in detail between the two measurement schemes. The positive and negative errorbars of $n_\textrm{dc}$ each indicate two standard deviations of the shown data at the $n=1$ plateau. The two data sets well agree within the experimental uncertainty. Averaging this data yields our best estimates for the average number of pumped electrons at the first plateau $n_\textrm{dc}=1.000\pm 0.006$ and $n_\textrm{count}=0.998\pm 0.004$ where we employ the 95\% uncertainty level.

Finally, the probability $P_1$ as a function of $V_\textrm{C2}$ and $V_\textrm{PL}$ is presented in Fig.~\ref{measured_plateau}(c). We observe that the robustness of the single-electron transfer with respect to $V_\textrm{PL}$ increases with decreasing $V_\textrm{C2}$. This phenomenon is due to an increase in the charging energy caused by a tightening of the electrostatic confinement of the pump dot. Here, we show this effect in the electron counting scheme as a consistency check of similar behaviour previously observed in the direct-current scheme \cite{Rossi2014,Seo2014}.

\section{Discussion}

In this work, we compare the direct current generated with a quantum dot pump with electron counting scheme at a relative uncertainty below a per cent. The main limiting factor of our experimental approach is the relatively low sensitivity of the charge detector. Typically, in order to confidently assess single-electron counting statistics, one has to trade between the size of the storage reservoir and the sensitivity of the sensor. Our device is designed to have a fairly large reservoir to minimize the back-action on the pumping mechanism. 

The disadvantage of this choice is the reduced performance of the readout. We estimate that the sensitivity of our detector is about~90~m$e$/$\sqrt{\textrm{Hz}}$. This indicates that it is possible to sense a single electron in about~8~ms of averaging time. However, we have chosen to integrate up to 750 ms between pumping pulses to reduce the uncertainty in the readout. In this context, the employed PID controller reduces the slow drifts in the SET current. Nevertheless, the limited reservoir-to-sensor capacitive coupling of about 0.005$e$ is mainly responsible for the non-ideal readout fidelity. The observed distributions of the signals for different numbers of pumped electrons are separated only by $3.2\sigma$. Hence, those counting events that fall further than~1.6$\sigma$ from the center of the distribution should be considered as misattributions in the most conservative scenario.
 
In the near future, we will integrate a metallic SET sensor next to the silicon reservoir. In this way, we estimate that the capacitive coupling and, hence, the sensitivity will be improved up to an order of magnitude. This will allow us to enhance the readout fidelity and reduce the counting uncertainty down to ppm levels, while keeping the back action on the pump insignificantly small.

Ultimately, a precise electron pump verified by error counting would, not only provide a supreme candidate for the realization of the quantum ampere \cite{Pekola2013}, but could also be harnessed in the quantum metrological triangle experiment \cite{Likharev1985,Milton2010} to test the fundamental constants of nature.

\section*{Acknowledgments}
We thank F.~Hudson, C. H. Yang, Y. Sun, G.~C.~Tettamanzi, I.~Iisakka, A.~Manninen, and A.~Kemppinen for useful discussions. We acknowledge financial support from the Australian Research Council (Grant No. DP120104710), the Academy of Finland (Grant No. 251748, 135794, 272806, 276528), Jenny and Antti Wihuri Foundation, The Finnish Cultural Foundation, and the Australian National Fabrication Facility. We acknowledge the provision of facilities and technical support by Aalto University at Micronova Nanofabrication Centre. A.R. acknowledges support from the University of New South Wales Early Career Research Grant~scheme.


\section*{References}
\bibliography{references}

\end{document}